\documentclass[pre,superscriptaddress,twocolumn]{revtex4}

\usepackage{graphicx}
\usepackage{amsmath}
\usepackage{amsfonts}
\usepackage{amssymb}

\begin{document}

\title{Diffusion of finite-size particles in channels with random walls}

\author{Maximilian Bauer}
\affiliation{Institute of Physics and Astronomy, University of Potsdam,
D-14476 Potsdam-Golm, Germany}
\affiliation{Physics Department, Technical University of Munich,
Garching, Germany}
\author{Alja\v{z} Godec}
\affiliation{Institute of Physics and Astronomy, University of Potsdam,
D-14476 Potsdam-Golm, Germany}
\affiliation{National Institute of Chemistry, Ljubljana, Slovenia}
\author{Ralf Metzler}
\email{rmetzler@uni-potsdam.de}
\affiliation{Institute of Physics and Astronomy, University of Potsdam,
D-14476 Potsdam-Golm, Germany}
\affiliation{Physics Department, Tampere University of Technology, FI-33101
Tampere, Finland}

\begin{abstract}
Diffusion of chemicals or tracer molecules through complex systems
containing irregularly shaped channels is important in many applications. Most
theoretical studies based on the famed Fick-Jacobs equation focus on the 
idealised case of infinitely small particles and reflecting boundaries.
In this study we use numerical simulations to consider the transport of 
\emph{finite-sized\/} particles through asymmetrical two-dimensional channels.
Additionally, we examine \emph{transient binding\/} of the molecules to the
channel walls by applying sticky boundary conditions. With the application of
diffusing pathogens in hydrogels in mind, we consider an ensemble of particles
diffusing in independent channels, which are characterised by common structural
parameters. We compare our results for the long-time effective diffusion 
coefficient with a recent theoretical formula obtained by Dagdug and Pineda 
[\textit{J. Chem. Phys.}, 2012, \textbf{137}, 024107].
\end{abstract}

\maketitle

\section{Introduction}
\label{sec:intro}

Civil aviation traffic has been broadly democratised within the last few decades,
leading to ever-increasing numbers of both business and pleasure passengers. The
downside of this increased worldwide connectivity is the very rapid global
spreading of diseases \cite{Brockmann2006Nature}. Simultaneously, new infectious
diseases are emerging, driven by human or ecologic causes \cite{daszak}, and
microorganisms are developing various forms of multiple drug resistance
\cite{antibiotics,sellar}. It is therefore imperative to develop more rapid,
mobile, and reliable methods for pathogen detection. One of the paths being
followed is the development of smart hydrogels, which respond to various
stimuli
\cite{Holtz1997Nature,Miyata1999Nature,Gawel2010Sensors,Wischerhoff2010SoMa}.

The scenario we have in mind is that viral pathogens diffuse into the hydrogel,
where they bind to specific sites and effect a local mechanical deformation. To
cause macroscopic effects, the pathogens' size must be comparable to the typical
mesh size of the hydrogel. A similar situation was investigated in a single
particle tracking study of submicron tracer beads in reconstituted F-actin
networks, observing \emph{anomalous diffusion\/} \cite{Wong2004PRL}
\begin{equation}
\label{eq:anomdiff}
\langle\mathbf{r}^2(t)\rangle\simeq t^\alpha,
\end{equation}
where the anomalous diffusion exponent $\alpha$ depended on the ratio of the tracer
bead size versus the typical mesh size of the actin network: for relatively large
beads $\alpha$ was shown to decrease to zero, while for small beads $\alpha$
converged to one, the case of normal diffusion \cite{Wong2004PRL}. Subdiffusion of
the type (\ref{eq:anomdiff}) was observed for similarly-sized tracer beads in
wormlike micellar solutions \cite{lene1}, as well as for the motion of various
objects in the macromolecularly crowded cytoplasm of living cells such as the
infectious pathway of adeno-associated viruses in living HeLa cells
\cite{Seisenberger2001Science}, as well as for submicron lipid and insulin
granules in living fission yeast and MIN6 insulinoma cells \cite{lene,tabei}. For
reviews on anomalous diffusion, see Refs.~\cite{report,report1}, and for
subdiffusion in
crowded systems, see Refs.~\cite{saxton,pt,franosch}.

Given these experimental findings it is pertinent to ask whether the motion of
pathogens in a hydrogel is equally subdiffusive. Here we address this question
by considering the path of the pathogen through the hydrogel as the motion of a
particle of finite size through a tortuous, corrugated channel with varying width. 
The particle gets repeatedly held up by bottlenecks in the channel, and may
transiently bind to the channel walls, representing the interaction with specific
binding site for the pathogen in the hydrogel. Using an ensemble of channel
geometries in our numerical analysis, we account for the different paths the
pathogen can take through the hydrogel. We find that indeed the particle in their
channel performs transient subdiffusion, that we analyse in terms of the anomalous
diffusion exponent, the number of successful moves with respect to the number of
simulations steps, and the effective long-time diffusivity, as function of the
characteristic channel geometry parameters.

The theoretical description of the motion of particles in confined geometries 
like channels (2D) or pores (3D) with varying width has a long-standing history 
(see \cite{Burada2009CPC} and references therein). In a seminal work Zwanzig
derived a modified Fick-Jacobs equation which is at the basis of most subsequent
quasi-one-dimensional descriptions \cite{Zwanzig1992JPC},
\begin{equation}
 \frac{\partial G(x,t)}{\partial t}=\frac{\partial}{\partial x}D(x)w(x)\frac{
\partial}{\partial x}\frac{G(x,t)}{w(x)}
\label{eq:modFJ}
\end{equation}
Here $G(x,t)$ describes the local concentration of particles at position $x$ and 
time $t$, $w(x)$ the width of the channel at position $x$ and most importantly 
$D(x)$ is an effective position-dependent diffusion coefficient. Subsequently,
several different forms for $D(x)$ were derived and applied to various systems 
\cite{Zwanzig1992JPC,Reguera2001PRE,Kalinay2005JCP,Kalinay2005PRE,Kalinay2006PRE,
Berezhkovskii2007JCP,Kalinay2008PRE,Martens2011PRE,Berezhkovskii2011JCP,
Kalinay2013PRE,Martens2013PRL}. Taking along only first order derivatives of the
width profile $w(x)$, Kalinay and Percus \cite{Kalinay2006PRE} (see also
Martens et al.~\cite{Martens2011PRE}) obtained the closed form result
\begin{equation}
D_{KP}(x)=D_0\frac{\arctan(w'(x)/2)}{w'(x)/2},
\label{eq:DKP}
\end{equation}
where $D_0$ denotes the position-independent diffusion coefficient in the absence
of confinement. We note that the presence of an $x$-dependent diffusivity in free
space is sufficient to effect various forms of anomalous diffusion \cite{andrey,
andrey1,fulinsky}.

However, the above forms of $D(x)$ are restricted to symmetric channels, i.e.,
channels with a straight centre-line. This constraint was removed in an approach
by Bradley \cite{Bradley2009PRE}, which was subsequently generalised by Dagdug
and Pineda to \cite{Dagdug2012JCP}
\begin{eqnarray} 
\nonumber
D_{DP}(x)&=&D_0\left(\frac{\arctan\left(y_0'(x)+\frac{w'(x)}{2}\right)}{w'(x)}
\right.\\
&&\left.-\frac{\arctan\left(y_0'(x)-\frac{w'(x)}{2}\right)}{w'(x)}\right),
\label{eq:DDP}
\end{eqnarray}
where $y_0(x)$ denotes the vertical position of the centre-line at horizontal
position $x$. Note that Eq.~(\ref{eq:DDP}) generalises all the previous results
for $D(x)$, for instance, one obtains the result (\ref{eq:DKP}) for symmetrical
channels by setting $y_0'(x)=0$ \cite{Dagdug2012JCP}.

As detailed by Zwanzig \cite{Zwanzig1992JPC}, in a system with periodic boundary 
conditions the effective diffusion coefficient in the long-time regime, $D_\mathrm{
eff}$, is obtained by using the following formula introduced by Lifson and Jackson
\cite{Lifson1962JCP} and generalised by Festa and Galleani d'Agliano
\cite{Festa1978PA},
\begin{equation}
\frac{1}{D_\mathrm{eff}}=\left\langle \frac{1}{D(x)w(x)}\right\rangle\langle 
w(x)\rangle,
\label{eq:deff}
\end{equation}
where $\langle\cdot\rangle$ denotes the average over one period. Thus, for any
channel with width profile $w(x)$ and centre-line $y_0(x)$, Eq.~(\ref{eq:DDP}) 
can be used to calculate $D_{DP}(x)$, which in turn is used in Eq.~(\ref{eq:deff})
to calculate the effective diffusion coefficient $D_\mathrm{eff,DP}$.

The above theories based on the Fick-Jacobs formalism apply to point-like particles,
that is, the particle can move along the channel as long as the width is not equal
to zero. For our scenario of pathogens moving in a channel, we argued that the
pathogen size is comparable to the channel width. Thus, the pathogen can only
fully cross the channel when the width profile $w(x)$ at any position is larger
than the particle size. Systems containing finite-size particles were indeed
studied in literature \cite{Dagdug2008JCP,Riefler2010JPCM}. Essentially, in all
formulae the channel width $w(x)$ has to be replaced by an effective channel width.

Another modification with respect to the Fick-Jacobs approach that we consider
here concerns the boundary conditions. Usually, reflecting boundary conditions
are used, i.e., when the particle collides with the channel walls, its perpendicular
motion is simply reversed. In the pathogen-hydrogel scenario, the pathogen will
(transiently) bind to specific binding sites incorporated into the hydrogel. Here
we explicitly include such effects by reactive boundary conditions, due to which
the model particle will transiently be immobilised by binding to the channel wall.
Moreover, it may immediately rebind to the channel wall after unbinding. As we will
see, this has a major effect on the particle motion.

Finally, the conventional Fick-Jacobs approach describes the motion of particles
in a single channel with quenched geometry. However, for the application to the
pathogen motion in the hydrogel, we mimic the possibility for particles to move
on different paths across the hydrogel by sampling an ensemble of tracer particles
in an ensemble of channel geometries. This ensemble of channels is characterised by
a common set of structural parameters.

The present study therefore represents an application of the Fick-Jacobs approach
to the biophysical problem of pathogen motion in a complex, confining environment.
At the same time, however, it significantly generalises the Fick-Jacobs model. The
paper is structured as follows. In the subsequent section we introduce the details
of the numerical approach used in this study. In section 3 we discuss how the
numerical results are analysed in terms of time and ensemble averaged observables.
In section 4 we present the detailed results. Section 5 puts our findings in
perspective with respect to theory by Dagdug and Pineda, before drawing our
conclusions in section 6.

\section{Simulation Details}
\label{sec:simproc}

\begin{figure}
\includegraphics[width=8cm]{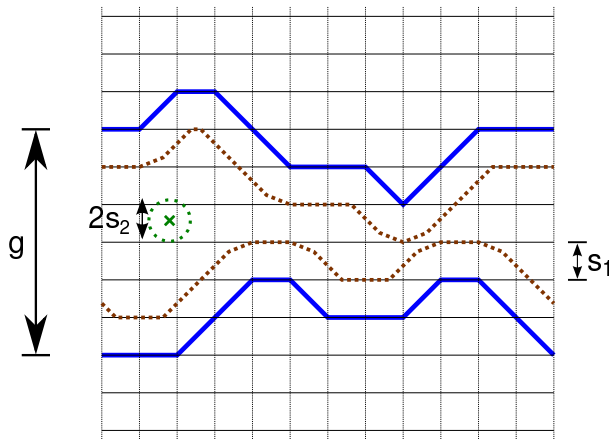}
\caption{Schematic of the channel with periodic boundary conditions. The blue
lines depict the channel walls, while the dotted (brown) lines mark the
excluded volume for a finite-sized particle. Parameters: gap opening $g=6$,
particle size $s_1=1$, step size $s_2=0.6$, ruggedness parameter $M_1=6$,
and lateral channel length $N=12$.}
\label{fig:scheme1}
\end{figure}

We study the diffusion through a two-dimensional channel with periodic boundary 
conditions in horizontal $x$-direction. In vertical $y$-direction the system is
limited by two walls, see Fig.~\ref{fig:scheme1}. These two walls are
described by $N$ points connected by straight lines, represented by the blue lines
in Fig.~\ref{fig:scheme1}. For numerical convenience the points of the channel
wall reside on a lattice with unit lattice constant, which effectively determines
the fundamental length scale of the system. Due to the horizontal periodic
boundary conditions the two leftmost and the two rightmost wall points are
identical. Their vertical distance (in $y$-direction) is denoted by $g$, see 
Fig.~\ref{fig:scheme1}. This is one of the fundamental parameters of the system
and is called gap opening parameter in the following.

We only consider channel configurations without touching or overlapping wall.
Moreover, we solely allow wall configurations in which the $y$-coordinates of
nearest neighbour points within one wall differ by at most unity. This excludes
the occurrence of extremely rugged walls. Thus the size of the system is
fully described by the gap opening parameter $g$ and the two `displacement vectors'
of size $(N-1)\times 1$ for the upper and lower walls. The $i^{\mathrm{th}}$ entry
of this displacement vector denotes the difference between the $y$-coordinate of
the $(i+1)^{\mathrm{st}}$ and $i^{\mathrm{th}}$ wall points for each of the two
walls ($i$ is counted from
left to right). Due to the constraints mentioned above only $0,\pm1$ are valid
entries, and the sum of all entries per wall must be zero to fulfil the periodic 
boundary requirement.

We consider channel walls with different contour lengths. The parameter $M_1$
describes the number of displacements of size $\pm1$ which occur in a wall.
We thus call it the `ruggedness parameter'.
Due to the periodic boundary constraint the number of jumps directed upwards 
must be equal to the one for jumps downwards for both walls. Consequently,
$M_1$ is an even number and lies in the interval $[0,N-1]$ (for odd $N$) or $[0,
N-2]$ (for even $N$). We only consider configurations in which $M_1$ is equal for
the upper and lower walls. However, this does not confine our study to symmetric
channels, compare Fig.~\ref{fig:scheme1}.

The position of the random walking particle is described by the position of its
centre of mass, illustrated by the green cross in Fig.~\ref{fig:scheme1}. Its
motion is off-lattice. This is shown in Fig.~\ref{fig:scheme1}, where a circle
of radius $s_2$ (the step size of the random walk) is drawn around the particle's
current position. For each step a random angle with respect to the $x$-axis is
drawn, and the particle attempts to move its centre to the corresponding point on
the dotted circle. To account for the diffusion of finite-sized particles through
the channel, before executing a step we check whether the distance from the current
position to the wall is sufficient. To this end, the minimal distance to the
wall is calculated for the trial position. Only if it is larger than the
particle size $s_1$, the step is actually performed. This accessible space is
limited by the two dotted brown lines in Fig.~\ref{fig:scheme1}. Their vertical
distance is the effective channel width for the finite-sized particle, and it is
the quantity to be inserted into Eqs.~(\ref{eq:DDP}) to (\ref{eq:deff}).
If the particle aims to move at a forbidden position, the step is cancelled
and the particle remains at its current position, but time is increased by one
unit. This corresponds to `sticky' boundary conditions, which mimic transient
binding to the channel wall.  

The diffusing particle is initially placed in the middle of the channel in both
$x$ and $y$ directions. However, if such a starting position is not possible in
the sense described above, the given channel configuration is dismissed and a
new one chosen. $T_{\mathrm{max}}$ random walk steps are performed and the
position in the $x$ direction is traced and analysed. If not stated otherwise,
for each parameter set $g$ and $M_1$ the results were averaged over $25,000$
configurations using the parameters $N=100$, $T_\mathrm{max}=10^6$, $s_1=1$, and
$s_2=0.6$.

\section{Evaluation procedure}
\label{sec:evalproc}

A quantity of central interest when tracking the motion of single particles is 
the time-averaged mean squared displacement (TA MSD)
\begin{equation} 
\overline{\delta_i^2(t)}=\frac{1}{T_\mathrm{max}-t}\int\limits_0^{T_\mathrm
{max}-t}dt'\Big[x_i(t'+t)-x_i(t')\Big]^2
\end{equation}
for the $i^\mathrm{th}$ time series $x_i(t)$ along the horizontal direction.
We use the fixed simulation time $T_\mathrm{max}$, and in what follows the 
bar denotes a time average. The TA MSD was subsequently averaged over all
configurations to obtain the ensemble \emph{and\/} time averaged mean squared
displacement (EATA MSD)
\begin{equation}
\left<\overline{\delta^2(t)}\right>=\frac{1}{N_\mathrm{conf}}\sum\limits_{i=1}^{
N_\mathrm{conf}}\overline{\delta_i^2(t)}.
\end{equation}
Thus, the usual ensemble-averaged MSD is nothing but a special case of the 
ensemble- and time-averaged MSD, where the point of reference is the starting 
position. We also consider the ensemble averaged mean squared displacement (EA
MSD) in $x$-direction,
\begin{equation}
\langle x^2(t)\rangle=\frac{1}{N_\mathrm{conf}}\sum\limits_{i=1}^{N_\mathrm{conf}}
\Big[x_i(t)-x_i(0)\Big]^2,
\end{equation}
where $N_\mathrm{conf}$ denotes the number of different configurations, in our
case $N_\mathrm{conf}=25,000$. Here and in the following $\langle\cdot\rangle$
denotes the ensemble average over channel realisations.

\begin{figure}
\includegraphics[width=8cm]{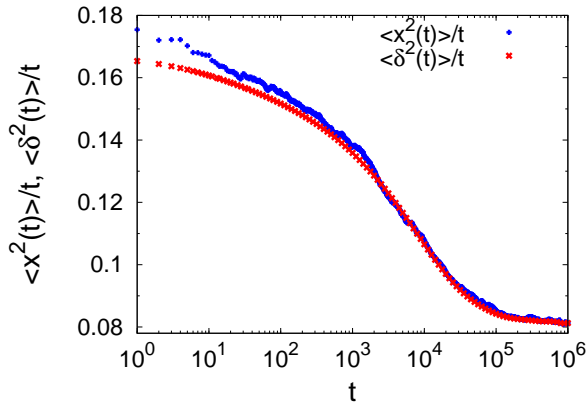}
\caption{Transient subdiffusion in an ensemble with gap opening parameter $g=6$
and wall ruggedness parameter $M_1=30$. We plot $\langle x^2\rangle/t$ (blue
symbols) and $\langle\overline{\delta^2}\rangle/t$ (red symbols) as function of
time $t$. Note the logarithmic abscissa.}
\label{fig:sample}
\end{figure}

As transient anomalous diffusion behaviour is not readily discernable in a
conventional log-log plot of the MSD versus time, we visualise the data in
terms of the MSD divided by time, as function of the
logarithm of time, see Refs.~\cite{Netz1995JCP,Saxton1996BPJ}. This method
emphasises deviations from normal diffusion behaviour: curves with a negative
slope represent subdiffusion. A typical plot is shown in Fig.~\ref{fig:sample}.
We observe a weaker form of subdiffusion for times from roughly $10$ to $10^3$
time steps. Around $10^3$ time steps, there is a turnover to a more pronounced
subdiffusive behaviour. This regime persists until some $10^5$ time steps, when
the terminal normal diffusive appears. We note that both time and ensemble MSD
coincide at longer times. On short time scales the EA MSD curve lies above the 
EATA MSD curve due to the fact that, by construction, at the beginning of each
trajectory the particle is placed in the middle of the channel. At such short
times the probability that the particle sticks to the wall is greatly
reduced compared to later times, and thus the EA MSD attains larger values than
the EATA MSD, which averages the behaviour along the entire time series. The
anomalous behaviour displayed in Fig.~\ref{fig:sample} is one of the major
results of this study.

In the following we study the slowing-down of the particle diffusion in terms of
two quantities. First, in the normal diffusive behaviour beyond $10^5$ time steps
we fit the EATA MSD data with a linear function in order to obtain the effective
long time diffusion coefficient $D_\mathrm{eff}$. This quantity is then compared
with the theoretical value $D_\mathrm{eff,DP}$ given by Eq.~(\ref{eq:deff}), since
for each channel configuration we calculate $D_{DP}(x)$ via Dagdug and Pineda's
formula (\ref{eq:DDP}), where $w(x)$ is replaced by the effective channel width.
We normalise the value of the effective long time diffusivity by the corresponding
value in absence of walls, $D_\mathrm{rel}=D_\mathrm{eff}/D_0$.
Second, we obtain the anomalous diffusion exponent $\alpha$ on time scales ranging
from $10^3$ to $10^5$ time steps. Similar results are obtained by studying the
mean maximal excursion of the particle \cite{Tejedor2010BPJ} (data not shown).

\section{Results}
\label{sec:results}

\subsection{Fixed channel wall configuration}
\label{sec:fixedbound}

Before studying the effect of different parameter sets $g$ and $M_1$ to
characterise the diffusion through this class of corrugated channels, we
investigate in detail the features seen in Fig.~\ref{fig:sample} from
simulations with fixed channel wall configurations.

\begin{figure}
\includegraphics[width=17.6cm]{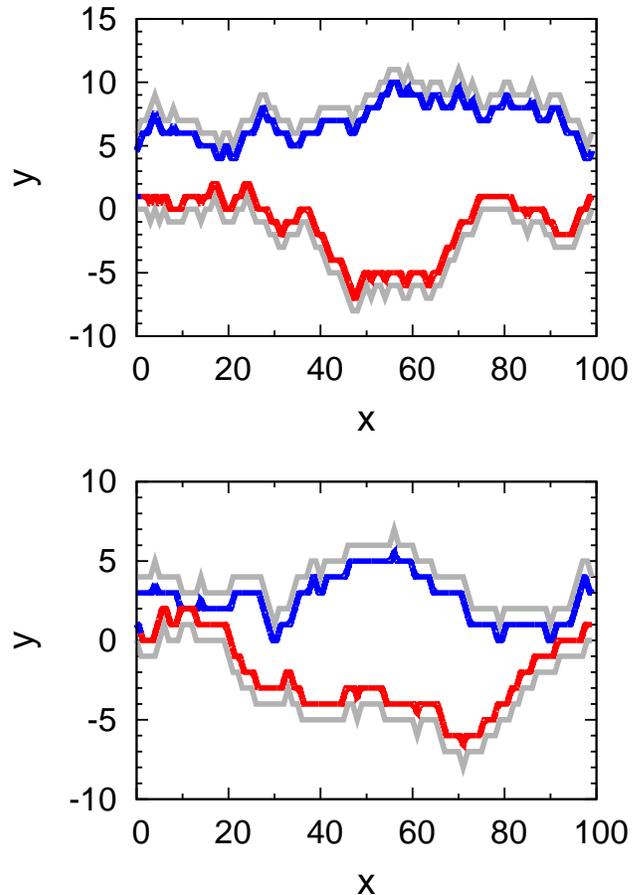}
\caption{Channel wall configurations characterised by the gap opening parameter 
$g=6$ and the wall ruggedness parameter $M_1=50$ (upper panel) and $g=4$, 
$M_1=30$ (lower panel). The actual upper and lower walls are plotted as grey lines.
The region bounded by the red and blue curves are accessible for the particle.}
\label{fig:Configs}
\end{figure}

We explicitly consider three exemplary configurations to illustrate the effect
of the sticky boundary conditions. These configurations are characterised by the
parameter pairs $g=6$ and $M_1=0$, $g=4$ and $M_1=30$, and $g=6$ and $M_1=50$.
The two configurations with non-flat walls are shown in Fig.~\ref{fig:Configs}.
The grey curves denote the actual position of the channel walls, while
the bold blue and red curves mark the region, which is inaccessible for the
particle's centre.

While for a given wall configuration by the choice of the gap opening parameter
$g$ we make sure that the particle finds sufficient space in the middle of the
channel where it is initially placed, it is a priori not certain that the particle
can traverse the entire channel. This is
the case when at some point the upper and lower walls overlap. Strictly speaking,
however, a passage is impossible only if there exists an overlap area whose width
is at last of the step size $s_2$. Otherwise, due to the finite step size the
particle can actually `tunnel' through such bottlenecks. The wall configuration
depicted in the upper panel of Fig.~\ref{fig:Configs} does not allow such a
tunnelling for the given parameters $g=6$ and $M_1=50$: the red and the black
curve do not overlap. The situation is different in the lower panel of
Fig.~\ref{fig:Configs} with $g=4$ and $M_1=30$: the channel is blocked for the
particle at $x\approx12$.

However, even if inaccessible regions in a given wall configuration exist but the
overlap of the walls stretches over less than the distance $2\times s_2$, such a
narrow straight
constitutes a severe entropic bottleneck for the diffusing particle: there exists
an appreciable possibility that the particle repeatedly sticks to the channel
walls. To quantify the influence of the sticky walls we binned the channels
into $99$ cells of length $1$ and extracted from our simulations how often the 
particle unsuccessfully tries to move to a new position while being in the 
corresponding bin.

\begin{figure}
\includegraphics[width=8.8cm]{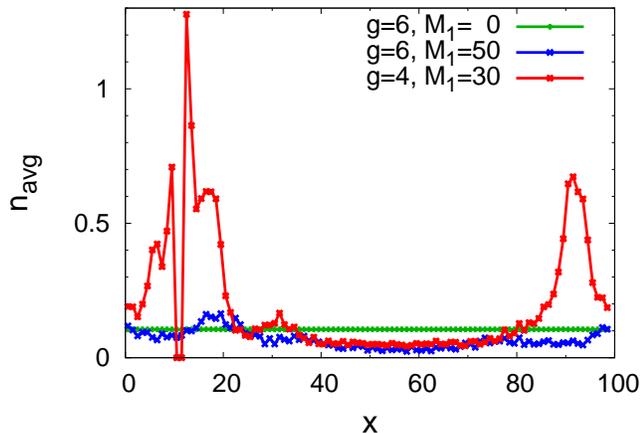}
\caption{Mean number $n_\mathrm{avg}$ of unsuccessful attempts to move to the
next position along the channel as a function of the position $x$ along the channel,
for the three wall configurations characterised by $g=6$ and $M_1=0$ (green
line), $g=6$ and $M_1=50$ (blue line), and $g=4$ and $M_1=30$ (red line).}
\label{fig:histogram}
\end{figure}

We first study the mean number $n_\mathrm{avg}$ of unsuccessful attempts for the
three above sample configurations as a function of the particle position along the
channel in Fig.~\ref{fig:histogram}. The green curve for the parameters $g=6$ and
$M_1=0$ shows that for flat walls $n_\mathrm{avg}$ is approximately constant.
The fact that this value is finite is a consequence of the sticky boundary 
condition at the edges of the system: namely, move attempts which would end at a 
position which is forbidden due to the finite size of the particle are not 
executed. If reflective walls were considered, any step could be executed 
and then $n_\mathrm{avg}=0$. In the present case, the value of $n_\mathrm{avg}$
depends on the step size and the (constant) width of the channel. From
Fig.~\ref{fig:histogram} we deduce that $n_\mathrm{avg}\approx0.1056$.

The blue curve in Fig.~\ref{fig:histogram} for $g=6$ and $M_1=50$ shows some
variation as function of $x$: where the channel is narrow, e.g., around $x\approx 
20$ in the upper panel of Fig.~\ref{fig:Configs}, $n_\mathrm{avg}$ is much higher
than at locations where the channel is wider, e.g., in the middle of the channel.
Comparing the minimum and maximum of $n_\mathrm{avg}$ along the channel, the
variation of $n_\mathrm{avg}$ makes up a factor of approximately $7$. This effect
is much more pronounced in the red curve in Fig.~\ref{fig:histogram} for the
parameters $g=4$ and $M_1=30$, corresponding to the lower panel of
Fig.~\ref{fig:Configs}: the curve is broken as two bins of the channel are
inaccessible for the particle. In the bin to the right of the channel blockage
the average number of unsuccessful tries is larger than $1$. In other regions of
the channel $n_\mathrm{avg}$ attains values similar to the ones in the other two
configurations. Hence, the mean number of unsuccessful motion attempts along the
channel directly reflects the effective channel width and thus the local transport
properties.

\begin{figure}
\includegraphics[width=8.8cm]{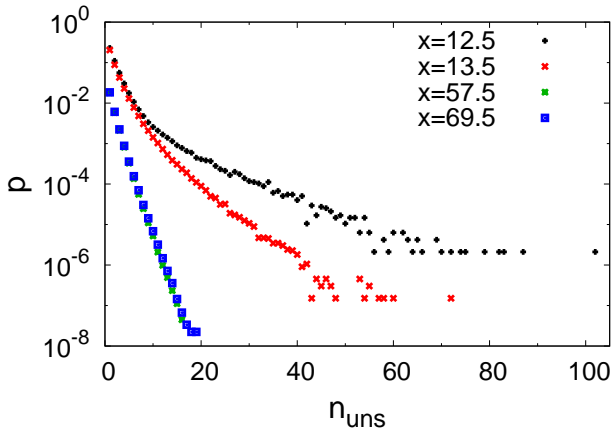}
\caption{Probability distribution for the number of subsequent unsuccessful motion
attempts, $n_\mathrm{uns}$, obtained with the wall configuration corresponding to
the lower panel of Fig.~\ref{fig:Configs} with parameters $g=4$ and $M_1=30$. We
show the statistics for the bins centred around: $x=12.5$ (black symbols), $x=13
.5$ (red symbols), $x=57.5$ (green symbols), and $x=69.5$ (blue symbols). The
symbols for $x=57.5$ are almost fully covered by those for $x=69.5$.}
\label{fig:extremedistrb}
\end{figure}

Additional information can be deduced from studying the probability distribution
$p(n_\mathrm{uns})$ of the number $n_\mathrm{uns}$ of unsuccessful attempts in a
row shown in Fig.~\ref{fig:extremedistrb}, where we focus on the most distinct
configuration with parameters $g=4$ and $M_1=50$ corresponding to the lower panel
of Fig.~\ref{fig:Configs}. For better visibility we only consider extreme cases:
namely, only the two bins with the highest and the two bins with the lowest mean
number of motion attempts. In all four cases, the probability to find higher values
of $n_\mathrm{uns}$ decreases. In regions where the channel is wide (green and
blue symbols in Fig.~\ref{fig:extremedistrb}) this decay is very fast, such that
within our simulation time there were never more than 19 subsequent unsuccessful
attempts. Otherwise, it becomes obvious that near severe bottlenecks the
distribution of waiting times is much more heavy-tailed (black and red symbols in 
Fig.~\ref{fig:extremedistrb}). Up to $100$ unsuccessful attempts in a row are
possible, with a probability of about $10^{-6}$.

With this information, let us go back to the features of Fig.~\ref{fig:sample}.
According to the Fick-Jacobs theory, whenever the width of the channel changes
in the form of a bottleneck or a bulge, this slows down the diffusion of the
particle \cite{Zwanzig1992JPC}: in the case of a bottleneck the particle may be
reflected back, while in the case of a bulge the particle may execute many
motion events off the minimal path along the channel. The effect of the entropic
bottlenecks in our present case is amplified by the presence of the sticky boundary
conditions. On all time-scales on which the particle interacts with the walls,
it is slowed down in comparison to a particle moving in free space. This induces
the transient subdiffusion in the ensemble and time averaged trajectories. On very
long time scales, when the configurations shown in Fig.~\ref{fig:Configs} are
equivalent to a single step size in a coarse-grained random walk on the whole
periodic structure, normal diffusive behaviour is restored, but now with a reduced
diffusion coefficient. This reduced coefficient takes into account all the
intermediate contacts with the channel walls. Thus, it is expected that more
corrugated and/or tighter channels, which imply more frequent interaction with the
walls should show reduced values of $\alpha$ and a reduced effective diffusion
coefficient on the ensemble level. To study these effects, in the following we
systematically study the impact of the parameters $g$ and $M_1$ on the transport
through the channels in ensembles of size 25,000.

\subsection{Simulations of channel ensembles}
\label{sec:simensmb}

\begin{figure}
\includegraphics[width=8.8cm]{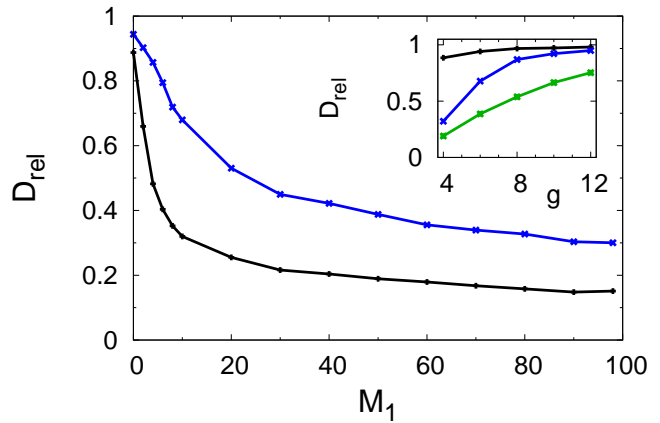}
\caption{Normalised effective long-time diffusion coefficient $D_\mathrm{rel}$ from
fitting of the simulations results, as function of the wall ruggedness
parameter $M_1$. Parameters: gap opening parameter $g=4$ (black symbols) and $g=6$
(blue symbols). Inset: $D_\mathrm{rel}$ as function of $g$ for $M_1=0$ (black
symbols), $M_1=10$ (blue symbols), and $M_1=50$ (green symbols).}
\label{fig:DofM1g}
\end{figure}

Two main simulations series were performed with the fixed values $g=6$ and $g=4$
for the gap opening parameter and 15 different values of the wall ruggedness
parameter $M_1$, spanning the whole possible range $[0,98]$. The fitted values of
the normalised effective diffusion coefficient in the long-time regime, $D_\mathrm{
rel}$, are depicted in Fig.~\ref{fig:DofM1g} as function of $M_1$. Here and in the
following, solely the EATA MSD values were used, as they supply the most extensive
data. The results obtained with the EA MSD data are very similar (data not shown), 
which is not surprising due to the ergodicity of the process at long times.
In both cases, an increase of the wall ruggedness (larger $M_1$ values) effects
slower effective diffusion. The slope of this decrease is steepest for small $M_1$
values and then gradually flattens off. Conversely, at fixed values of $M_1$ the
effective diffusion is always substantially faster for $g=6$ than $g=4$.

To study the impact of the gap opening parameter $g$ in more detail, we took five
different values at fixed values of the wall ruggedness parameter, namely, $M_1=0$,
$M_1=10$, and $M_1=50$. Thus, we consider flat channels, slightly corrugated, and
heavily corrugated channels. The corresponding results for the fitted values of
the normalised effective diffusion constant $D_\mathrm{rel}$ are shown in the
inset of Fig.~\ref{fig:DofM1g}. For fixed value of $g$ we see once more that the
diffusion is quickest when the wall is smoother or, i.e., when $M_1$ is smaller.
As was already observed in the preceding paragraph a wider gap opening at a 
fixed value of $M_1$ leads to faster diffusion. Thus, wider channels can be
traversed quicker.

\begin{figure}
\includegraphics[width=8.8cm]{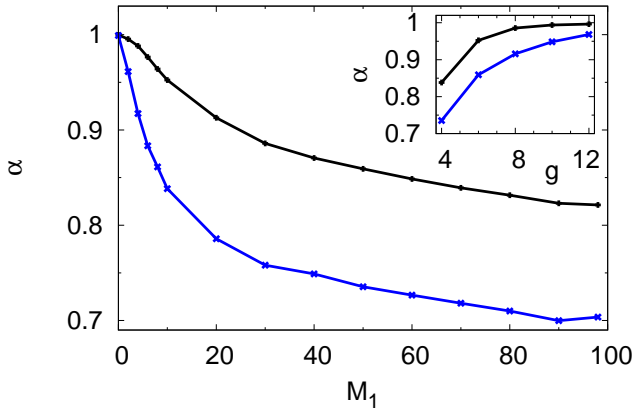}
\caption{Anomalous diffusion exponent $\alpha$ as function of wall ruggedness
$M_1$ from power-law fit of the EATA MSD data. Parameters: $g=4$ (blue) and 
$g=6$ (black). Inset: $\alpha$ as function of $g$ for $M_1=10$ (black), and
$M_1=50$ (blue).}
\label{fig:alphaofM1g}
\end{figure}

At first sight surprisingly, we observe that even for completely flat upper and 
lower channel walls ($M_1=0$, black line in the inset of Fig.~\ref{fig:DofM1g})
the diffusion in tighter channels is slowed down in comparison to the situation
in free space. This is not expected in systems with reflecting boundaries, which
are usually described with the Fick-Jacobs equation. However, for the finite-size
particles studied here it is the result of the sticky boundary conditions at the
channel walls: in a tighter channel the particle is more often close to the walls
and binds transiently. Alternatively, the slow-down due to the interaction with
the walls can be quantified by measuring the anomalous diffusion exponent $\alpha$
in the intermediate time regime, on time scales $10^3$ to $10^5$ simulations steps.
The fitted values for $\alpha$ are depicted in Fig.~\ref{fig:alphaofM1g} as
function of the ruggedness $M_1$. The same trend as for the effective diffusion
coefficient is seen for the anomalous diffusion exponent $\alpha$ of the transient 
subdiffusive regime. For increasing contour lengths of the channel wall the motion
is increasingly subdiffusive. As expected, the transient subdiffusion is heavier
for the tighter channel ($g=4$).

The dependence of $\alpha$ on the gap opening parameter is shown in the inset 
of Fig.~\ref{fig:alphaofM1g}. In this case, only values obtained with corrugated
channels ($M_1=10$ and $M_1=50$) are shown.\footnote{In flat channels with $M_1=0$
no transient subdiffusion occurs, see above.} Again, the curves for $\alpha$ are
similar to the ones obtained for the effective diffusion coefficient: $\alpha$ is
an increasing function of $g$ and $M_1=10$ leads to less pronounced transient
subdiffusion than $M_1=50$. This analogy motivates the study of the relation
between $\alpha$ and $D_\mathrm{rel}$ in more detail.

\begin{figure}
\includegraphics[width=8.8cm]{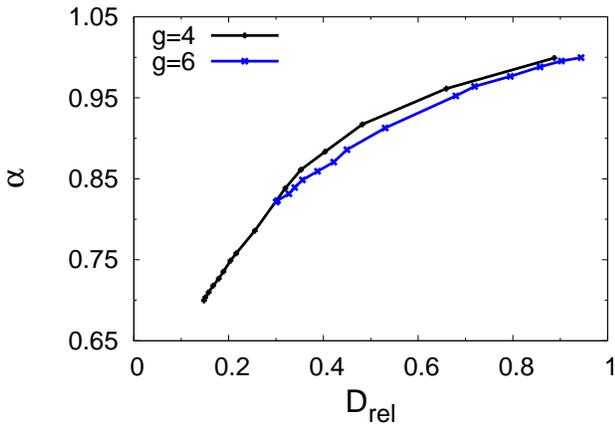}
\caption{Anomalous diffusion exponent $\alpha$ of the transiently subdiffusive
regime as function of the long-time diffusion exponent $D_\mathrm{rel}$ with
gap opening parameters $g=4$ (black) and $g=6$ (blue).}
\label{fig:corlalphadeff}
\end{figure}

In Fig.~\ref{fig:corlalphadeff} we plot all $\alpha$ values for gap opening $g=4$
and $g=6$ as function of the corresponding fitted values of $D_\mathrm{rel}$. 
The results show that there is a strong (nonlinear) correlation between both 
parameters. For increasing values of $D_\mathrm{rel}$ the value of $\alpha$ also
increases, with decreasing slope. The relation between both parameters is bijective,
thus, both quantities are appropriate and sufficient to quantify the slow-down of
the motion along the channel. For similar values of $D_\mathrm{rel}$ the values for
$\alpha$ obtained with the tighter channels ($g=4$) are slightly larger than those
for the wider channels ($g=6$). However, this fact should not be overstated: all
data sets were fitted in the time interval $10^3$ to $10^5$, irrespective of the
exact shapes of the curves. It is conceivable that a closer connection between
the values for $\alpha$ could have been obtained by choosing the fitting time
window for each curve individually.

\begin{figure}
\includegraphics[width=8.8cm]{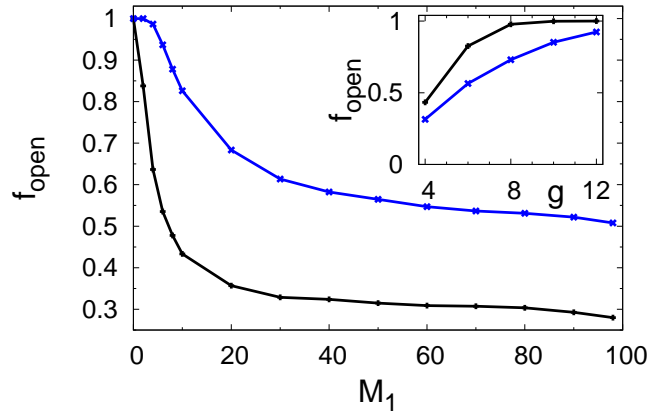}
\caption{Fraction $f_\mathrm{open}$ of unblocked configurations as function of
the channel ruggedness $M_1$ for channel opening $g=4$ (black) and $g=6$ (blue).
Inset: $f_\mathrm{open}$ as function of $g$ for $M_1=10$ (black) and $M_1=50$
(blue).}
\label{fig:fopenofM1g}
\end{figure}

As mentioned above, in an ensemble of systems with corrugated boundaries not all 
channels can be traversed completely. If a channel is blocked somewhere, the
corresponding squared displacement of the particle position has an upper limit.
On an ensemble level these trajectories will reduce the average values of 
$\alpha$ and $D_\mathrm{rel}$. Thus, it is important to extract from our
simulations solely the unblocked configurations. The corresponding parameter
$f_\mathrm{open}$ is plotted as function of the ruggedness $M_1$ for fixed gap
openings $g=4$ (black symbols) and $g=6$ (blue symbols) in 
Fig.~\ref{fig:fopenofM1g}, and as function of the gap opening $g$ (for $M_1=10$
and $M_1=50$) in the inset of Fig.~\ref{fig:fopenofM1g}. An inspection of
Fig.~\ref{fig:fopenofM1g} shows that this fraction is a decreasing function of
$M_1$ and an increasing function of $g$. Overall, the curves look similar to the
ones of the normalised effective diffusion coefficient $D_\mathrm{rel}$ (compare 
Fig.~\ref{fig:DofM1g}).

\begin{figure}
\includegraphics[width=8.8cm]{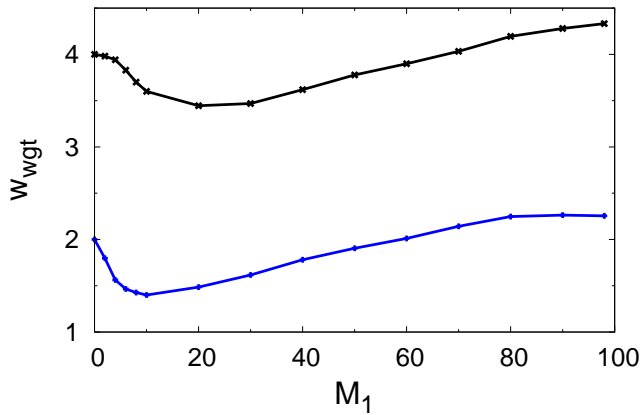}
\caption{Weighted effective channel width $w_\mathrm{wgt}$ as function of channel
ruggedness $M_1$ for gap opening $g=6$ (black line and symbols) and $g=4$ (blue
line and symbols).}
\label{fig:wwgtofm1}
\end{figure}

To better understand why above a certain threshold of the boundary ruggedness 
parameter $M_1$ the effective diffusion coefficient only decreases slightly 
(see Fig.~\ref{fig:DofM1g}), it is instructive to study the average weighted
effective width $w_\mathrm{wgt}$ of the channels. Here, weighted means that for 
any blocked channel the width is set to zero. The weighted width is plotted as
function of $M_1$ in Fig.~\ref{fig:wwgtofm1}. For increasing, yet small values
of $M_1$ the effective weighted channel width decreases, until it reaches a
shallow minimum beyond which $w_\mathrm{wgt}$ levels off to a plateau. While
heavily rugged channel walls, in principle, can be tighter or more spacious 
than a flat configuration with the same gap opening $g$, the tighter ones are in 
constant `danger' of precluding the particle passage. Thus, the average width of
those traversable channels is larger for more rugged wall configurations.
This facilitates the transport through these configurations, as the sticky 
boundaries are further away. However, as remarked earlier, more rugged walls
slow down the diffusion due to the occurrence of bulges and constrictions
\cite{Zwanzig1992JPC}, such that we have two opposing effects, which mostly
(almost) cancel each other. Consequently, $w_\mathrm{wgt}$ (and thus $D_\mathrm{
rel}$) reaches a plateau above a threshold value of $M_1$.

\begin{figure}
\includegraphics[width=8.8cm]{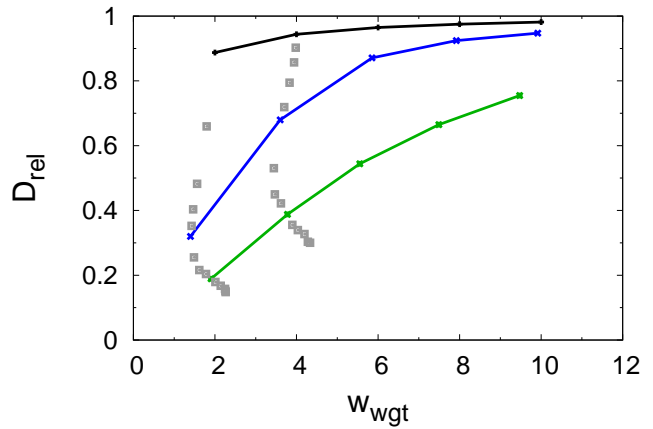}
\caption{Normalised effective diffusion coefficient $D_\mathrm{rel}$ as function
of the weighted effective channel width $w_\mathrm{wgt}$ for $M_1=0$ (black),
$M_1=10$ (blue line), $M_1=50$ (green line) and other values of $M_1$ (grey
symbols).}
\label{fig:Drelwwgt}
\end{figure}

Fig.~\ref{fig:Drelwwgt} shows the normalised effective diffusion coefficient $D_
\mathrm{rel}$ as function of the weighted channel width $w_\mathrm{wgt}$. For
better visibility data points with identical $M_1$ values are connected by lines
($M_1=0$: black line, $M_1=10$: blue line, and $M_1=50$: green line). While for
a fixed value of $M_1$ more spacious channels allow faster diffusion, the heavy
scatter between values of $D_\mathrm{rel}$ obtained with similar values of $w_
\mathrm{wgt}$ (grey symbols) shows that the knowledge of the mean channel width
of an ensemble of channels alone is insufficient to predict the transport
properties. 

\section{Comparison with Dagdug and Pineda's formula}
\label{sec:compdagdug}

\begin{figure}
\includegraphics[width=17.2cm]{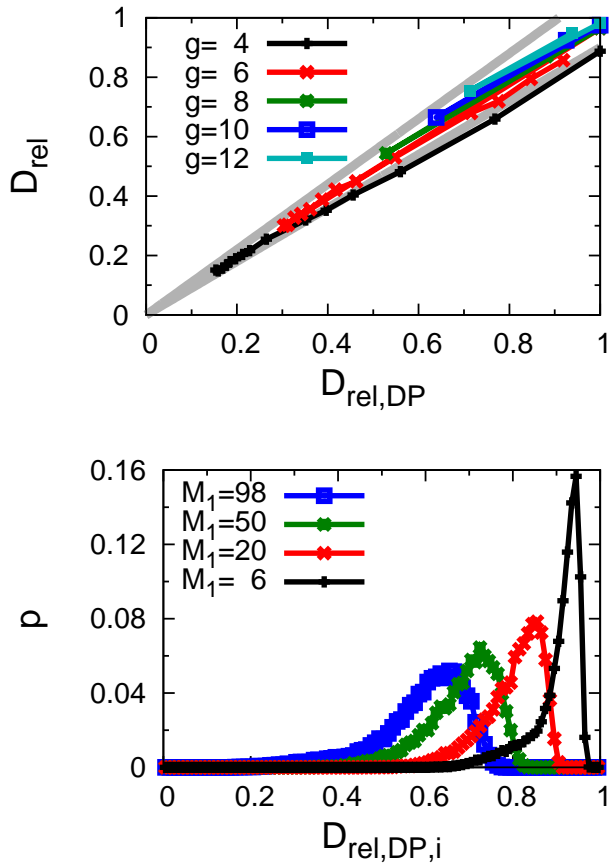}
\caption{Upper panel: normalised effective diffusion coefficient $D_\mathrm{rel}$
from our simulations in the sticky, corrugated channel as function of the value
$D_\mathrm{rel,DP}$ obtained from the result (\ref{eq:DDP}) of Dagdug and Pineda.
Black: gap opening parameter $g=4$, red: $g=6$, green: $g=8$, blue: $g=10$, and
cyan: $g=12$. The two grey lines mark the range of $\pm 10\%$ around the
theoretical value. Lower panel: distributions of effective diffusion coefficients
for individual unblocked channel configurations obtained with Dagdug and Pineda's
formula. Parameters: $g=6$ in all four cases. $M_1=6$: black, $M_1=20$: red,
$M_1=50$: green, $M_1=98$: blue.}
\label{fig:Drelcompdistr}
\end{figure}

In order to compare our results obtained from ensembles of channel configurations
with the result of Dagdug and Pineda, we make a simplifying assumption: for all
unblocked configurations, we determine $D_{DP}(x)$ from Eq.~\ref{eq:DDP} and
subsequently $D_\mathrm{eff,DP}$ from Eq.~\ref{eq:deff}. For all blocked
configurations the effective diffusivity $D_\mathrm{eff,DP}=0$ accounts for the 
fact that on long time-scales particles in these configurations do not contribute
significantly to the MSD. Finally, we average over all configurations in the
ensemble and normalise through division by the diffusion coefficient in free
space, $D_\mathrm{rel,DP}=\langle D_\mathrm{eff,DP}\rangle/D_0$. We compare these
values with our fitted values of the normalised diffusion coefficient in the upper
panel of Fig.~\ref{fig:Drelcompdistr}, where data points with the same value of the
gap opening $g$ are represented in the same colour.

This is the central result of our study: As demonstrated in the upper panel of
Fig.~\ref{fig:Drelcompdistr}, for fixed values of the gap opening $g$ there is
a linear relation between $D_\mathrm{rel}$ and $D_\mathrm{rel,DP}$. Most of the
data points are located within a $10\%$ confidence interval around Dagdug and
Pineda's value. More explicitly, a linear fit of the data points obtained in a
simulation series yields the following results. For $g=4$, the fitted relation
between the two is $D_\mathrm{rel}=(0.018\pm0.004)+(0.853\pm0.009)D_\mathrm{
rel,DP}$, while for $g=6$ we find $D_\mathrm{rel}=(0.032\pm0.095)+(0.903\pm0.007)
D_\mathrm{rel,DP}$. For $g=8$, $10$, and $12$ we did not fit the data as there
were only three values available.

The fact that the slope of the fits is somewhat below $1$ shows that Dagdug and 
Pineda's formula, which only applies to systems with perfectly reflecting
boundaries slightly overestimates the diffusion coefficient compared to our system
with sticky boundary conditions. Thus, as expected the additional interaction with
the boundaries further slows down the diffusion, which is already reduced by the
occurrence of entropic bottlenecks. This reasoning is further substantiated by
the observation that in tighter channels (with $g=4$), where these surface effects
play a more important role, the slope of the conversion formula is smaller, and
the deviation from Dagdug and Pineda's formula is more pronounced. Given the quite
intricate form of the effective channel width (see Figs.~\ref{fig:scheme1} and
\ref{fig:Configs}) it is not feasible to quantify this effect analytically. However,
in the future other values of $s_1$ and $s_2$ could be considered to study the
magnitude of the correction terms numerically.

That deviations of our results from Dagdug and Pineda's formula are also based on
the fact that their analysis applies to one given channel configuration. Driven by
the physical application we here consider an ensemble of different channel wall
configurations, solely defined by the fixed macroscopic parameters $g$ and $M_1$.
Individual configurations may therefore differ considerably. This is illustrated
in the lower panel of Fig.~\ref{fig:Drelcompdistr}, where for all unblocked
configurations the expected effective diffusion coefficient was calculated with
Dagdug and Pineda's formula. For $g=6$ and four different values of $M_1$ we see
that the distribution of values of $D_\mathrm{rel,DP}$ is far from narrow. The
values increasingly scatter for more rugged conformations (higher values of $M_1$).
Thus, the semi-quantitative agreement of our data with the theoretical prediction
on an ensemble level is indeed remarkable.

\section{Conclusion and outlook}
\label{sec:concloutl}

We studied the motion of finite-size particles through randomly corrugated
channels with sticky walls, observing transient anomalous diffusion of the
particles in their passage of the channel. We also obtained the long-time
diffusion coefficient for this motion on time scales over which normal
Brownian diffusion is restored. The control parameters in our study were the
gap opening parameter fixing the distance between the channel walls at the
entrance and exit of the channel, as well as the wall ruggedness parameter
setting the maximal variation of the channel wall configuration. We quantified
the dependence of the anomalous diffusion exponent and long-time diffusion
coefficient as function of the ruggedness and gap opening, and showed that
both quantities are in fact correlated. We especially analysed the blocked
channels, which the particle cannot fully traverse. The long-time effective
diffusion coefficient was shown to agree well with the prediction for point-like
particles in channels with reflecting boundary conditions by Dagdug and Pineda.
Translated into the language of pathogens in hydrogels, whose motion we want
to mimic in our numerical study of an ensemble of `parallel' channels with
identical gap opening and ruggedness, our study provides statistical information
on how many channels are blocked for the pathogen passage.

In future studies it might be interesting to replace our minimal model of 
interactions with the constituents of the hydrogel with a more realistic model.
In particular, the introduction of dynamic boundaries, for example by varying 
the gap opening during the simulations might be worth considering in order to 
model the structural change of the hydrogel due to the binding of pathogens.

MB and AG acknowledge funding from the German Federal Ministry for Education
and research, and RM from the Academy of Finland within the FiDiPro scheme.

\end{document}